\documentclass[usenatbib]{mn2e}
\usepackage{epsfig}
\usepackage{amssymb}
\usepackage{lscape,graphicx}
\usepackage{longtable}
\usepackage{psfig}

\usepackage{lscape}

\usepackage{rotating}

\usepackage{graphicx}
\usepackage{epstopdf}

\def\gsim{\mathrel{\raise0.35ex\hbox{$\scriptstyle >$}\kern-0.6em
\lower0.40ex\hbox{{$\scriptstyle \sim$}}}}
\def\lsim{\mathrel{\raise0.35ex\hbox{$\scriptstyle <$}\kern-0.6em
\lower0.40ex\hbox{{$\scriptstyle \sim$}}}}

\begin{document}
\vspace{-5cm}
\title[An SMG at $z$\,=\,4.76 in the ECDFS]
{\vspace{-1.cm}A submillimetre galaxy at $z$\,=\,4.76 in the LABOCA survey of the Extended Chandra Deep Field South}

\author[Coppin et al.]{ 
\parbox[t]{\textwidth}{\vspace{-1cm}
K.\,E.\,K.\ Coppin,$^{\! 1}$ Ian Smail,$^{\! 1}$ D.\,M.\
Alexander,$^{\! 2}$ A.\ Weiss,$^{\! 3}$ F.\ Walter,$^{\! 4}$ A.\,M.\
Swinbank,$^{\! 1}$ T.\,R.\ Greve,$^{\! 4}$ A.\ Kovacs,$^{\! 3}$  C.\
De Breuck,$^{\! 5}$ M.\ Dickinson,$^{\! 6}$ E.\ Ibar,$^{\! 7}$
R.\,J.\ Ivison,$^{\! 7,8}$ N.\ Reddy,$^{\! 6}$ H.\ Spinrad,$^{\! 9}$
D.\ Stern,$^{\! 10}$ W.\,N.\ Brandt,$^{\! 11}$ S.\,C.\ Chapman,$^{\!
12}$ H.\ Dannerbauer,$^{\! 4}$  P.\ van Dokkum,$^{\! 14}$ J.\,S.\
Dunlop,$^{\! 7}$ D.\ Frayer$^{\! 13}$ E.\ Gawiser,$^{\! 15}$ J.\,E.\
Geach,$^{\! 1}$ M.\ Huynh,$^{\! 13}$  K.\,K.\ Knudsen,$^{\! 16}$
A.\,M.\ Koekemoer,$^{\! 17}$ B.\,D.\ Lehmer,$^{\! 2}$ K.\,M.\ Menten,$^{\! 3}$ C.\ Papovich,$^{\! 18}$  H.-W.\ Rix,$^{\! 4}$ E.\ Schinnerer,$^{\! 4}$  J.\,L.\ Wardlow,$^{\! 2}$ P.\,P.\ van der Werf$^{19}$}\\\\
$^{1}$ Institute for Computational Cosmology, Durham University, South Road, Durham, DH1 3LE, UK\\
$^{2}$ Department of Physics, Durham University, South Road, Durham, DH1 3LE, UK \\
$^{3}$ Max-Planck-Institut f\"{u}r Radioastronomie, Auf dem H\"{u}gel 69, Bonn, D-53121, Germany \\
$^{4}$ Max-Planck-Institut f\"{u}r Astronomie, K\"{o}nigstuhl 17, Heidelberg, D-69117, Germany\\
$^{5}$ European Southern Observatory, Karl-Schwarzschild Strasse, 85748 Garching bei M\"{u}nchen, Germany\\
$^{6}$ National Optical Astronomy Observatory, P.O.\ Box 26732, Tucson, AZ 85726, USA \\
$^{7}$ SUPA\thanks{Scottish University Physics Alliance}, Institute for Astronomy, University of Edinburgh, Royal Observatory, Blackford Hill, Edinburgh, EH9 3HJ, UK\\
$^{8}$ UK Astronomy Technology Centre, Royal Observatory, Blackford Hill, Edinburgh, EH9 3HJ, UK \\
$^{9}$ Department of Astronomy, University of California at Berkeley, Mail Code 3411, Berkeley, CA 94720, USA \\
$^{10}$ Jet Propulsion Laboratory, California Institute of Technology, Mail Stop 169-527, Pasadena, CA 91109, USA \\
$^{11}$ Department of Astronomy and Astrophysics, 525 Davey Lab, Pennsylvania State University, University Park, PA 16802, USA\\
$^{12}$ Institute of Astronomy, Madingley Road, Cambridge, CB3 0HA, UK \\
$^{13}$ Infrared Processing and Analysis Center, MS220-6, California Institute of Technology, Pasadena, CA 91125, USA\\
$^{14}$ Department of Astronomy, Yale University, PO Box 208101, New Haven, CT 06520 USA \\
$^{15}$ Physics \& Astronomy Department, Rutgers University, Piscataway, NJ 08854, USA \\
$^{16}$ Argelander Institute for Astronomy, University of Bonn, Auf dem H\"{u}gel 71, D-53121 Bonn, Germany \\
$^{17}$ Space Telescope Science Institute, 3700 San Martin Drive, Baltimore, MD 21218, USA \\
$^{18}$ G.P.\ \& C.M.\ Mitchell Institute for Fundamental Physics, Department of Physics, Texas A\&M University, College Station, TX 77843, USA \\
$^{19}$ Leiden Observatory, Leiden University, PO Box 9513, NL-2300 RA Leiden, the Netherlands
}
\maketitle
%
%
%
\begin{abstract}
We report on the identification of the highest redshift submillimetre-selected source currently known: LESS\,J033229.4$-$275619.  This source was detected in the Large Apex BOlometer CAmera (LABOCA) Extended Chandra Deep Field South (ECDFS) Submillimetre Survey (LESS), a sensitive 870-$\mu$m survey ($\sigma_{870\mu\rm m} \sim 1.2$\,mJy) of the full $30'\times 30'$ ECDFS with the LABOCA camera on the Atacama Pathfinder EXperiment (APEX) telescope.  The submillimetre emission is identified with a radio counterpart for which optical spectroscopy provides a redshift of $z=4.76$.  We show that the  bolometric emission is dominated by a starburst with a star formation rate of $\sim 1000$\,$M_\odot$\,yr$^{-1}$, although we also identify a moderate luminosity Active Galactic Nucleus (AGN) in this galaxy. Thus it has characteristics similar to those of $z\sim 2$ submillimetre galaxies (SMGs), with a mix of starburst and obscured AGN signatures.  This demonstrates that ultraluminous starburst activity is not just restricted to the hosts of the most luminous (and hence rare) QSOs at $z\sim 5$, but was also occurring in less extreme galaxies at a time when the Universe was less than 10 per cent of its current age.  Assuming that we are seeing the major phase of star formation in this galaxy, then we demonstrate that it would be identified as a luminous distant red galaxy at $z\sim 3$ and that the current estimate of the space density of $z>4$ SMGs is only sufficient to produce $\gsim10$ per cent of the luminous red galaxy population at these early times.  However, this leaves open the possibility that some of these galaxies formed through less intense, but more extended star formation events.  If the progenitors of all of the luminous red galaxies at $z\sim3$ go through an ultraluminous starburst at $z\gsim4$ then the required volume density of $z>4$ SMGs will exceed that predicted by current galaxy formation models by more than an order of magnitude.
\end{abstract}


\section{Introduction}

It is now well established that a significant population of luminous red galaxies exist out to at least  $z\sim1.5$--2 and it has been claimed that most of these are massive and that some of them are old ($>10^{11}$\,M$_{\odot}$, $\gsim 2$\,Gyr) galaxies (\citealt{Cimatti04}; \citealt{Daddi05}; \citealt{Kong06}; \citealt{McGrath07}; \citealt{Hartley08}).   There are indications that many of these massive galaxies were largely formed at earlier times, $z\sim 2.5$--3.5, with a fraction of them already having red colours, suggesting an aging stellar population
(\citealt{Marchesini07}; \citealt{Marchesini08}). If correct, this suggests that these galaxies must have formed the bulk of their stellar populations at $z>4$ (\citealt{Daddi05}; \citealt{Cimatti08}; \citealt{Stockton08}). 
The implied star formation rates (SFRs) for these progenitors would then be $\sim 10^2$--10$^3$\,M$_\odot$\,yr$^{-1}$ and if such vigorous activity occurred in a single halo, it is likely that these systems would appear as ultraluminous infrared galaxies (ULIRGs), whose redshifted far-infrared emission would make them bright submillimetre galaxies (SMGs). However, the confirmation of large numbers of galaxies forming stars at a rate of $\sim 10^3$\,M$_\odot$\,yr$^{-1}$  at such high redshifts could pose a serious challenge to current hierarchical models, which  predict a $z>4$ SMG surface density about an order of magnitude less ($\sim10$\,deg$^{-2}$) than would be implied if all massive high-redshift galaxies form through this process (e.g.\ \citealt{Granato04}; \citealt{Baugh05}; \citealt{Hopkins05}; \citealt{Bower06}; \citealt{Swinbank08}).

Due to the negative K-correction, SMGs can essentially be detected out to $z\sim8$ (see \citealt{Blain02}), but the confirmation of SMGs (if they exist) at these redshifts poses a challenge.  The large beamsize of submm cameras (e.g.\ full width at half maximum (FWHM) $\sim 19''$ for LABOCA) precludes the determination of precise positions for large samples of SMGs without very significant investment in sub-/millimetre interferometry (e.g.\ \citealt{Downes99}; \citealt{Gear00}; \citealt{Dannerbauer02,Dannerbauer08}; \citealt{Wang07}; \citealt{Younger07,Younger08}).  This makes determining their nature through multiwavelength avenues extremely challenging, since there are typically many objects within this large beam that could be the source of the submillimetre emission.  The most successful route to identifying the counterparts of SMGs has been to search for 1.4-GHz counterparts (e.g.\ \citealt{Ivison02}), exploiting the well-known far-infrared--radio correlation (\citealt{Condon92}; \citealt{Helou93}).  With deep, $\lsim 10\,\mu$Jy rms, radio maps this provides precise positions for about 60 per cent of the SMG population ($S_{850}\gsim4$\,mJy), allowing follow-up spectroscopy and yielding a redshift distribution for radio-identified SMGs peaking at $z\sim2.2$ \citep{Chapman05}.  With the advent of {\it Spitzer}, a similar approach has exploited 24\,$\mu$m counterparts to SMGs, when radio counterparts were unavailable, increasing the fraction of identified SMGs to $\gsim 70$ per cent (e.g.\ \citealt{Ivison04, Ivison07}; \citealt{Pope06}).  Unfortunately the radio and mid-infrared emission dim significantly with redshift, which means that the question of whether a significant proportion of SMGs lie at $z\gsim 4$ remains  open.

Luminous submillimetre-emitting sources are of course known at $z\sim 4$--6 (e.g.\ \citealt{McMahon94}; \citealt{Omont96}; \citealt{Ivison98b}; \citealt{Carilliqso,Carilli01}; \citealt{Archibald01};  \citealt{Priddey08}), but these are all associated with rare and extreme Active Galactic Nuclei (AGN) whose space densities are so low that they are probably not a significant progenitor population for the luminous red galaxies seen at $z\sim 3$ (although this conclusion depends upon the assumed duty cycle of the AGN).  In contrast, the largest current redshift survey of the more numerous submillimetre-selected galaxy population \citep{Chapman05} identified no SMGs at $z\gsim4$. As noted above, the need for precise locations for the SMGs from their $\mu$Jy radio emission biased the redshift distribution to $z\lsim 3.5$ (\citealt{Chapman05}), and there could be a high-redshift tail (\citealt{Dannerbauer02, Dannerbauer08}; \citealt{Wang07}; \citealt{Younger07}; \citealt{Greve08}; \citealt{Wang09}; \citealt{Knudsen09}).  \citet{Pope06} and \citet{Aretxaga07} derived photometric redshift estimates (albeit much less precise than spectroscopic redshifts) for possible counterparts to SMGs and proposed that a relatively low fraction ($<10$ per cent) of SMGs lie at $z\gsim4$, suggesting that the \citet{Chapman05} redshift distribution is representative of all SMGs.  Nevertheless, there has been recent confirmation of three radio-identified SMGs lying at $z=4$--4.5 (\citealt{Daddi08}; \citealt{Capak08}; \citealt{Schinnerer08}), indicating that a tail of SMGs does exist at $z\gsim4$.

In this paper we focus on a luminous SMG discovered in our deep 870-$\mu$m Large Apex BOlometer CAmera (LABOCA) Extended Chandra Deep Field South (ECDFS) Survey (LESS; PIs Smail, Walter \& Weiss; Weiss et al.\ in preparation, hereafter W09) which is associated with a faint radio and mid-infrared source. Spectroscopic and photometric observations identify this source as an ultraluminous far-infrared galaxy at $z=4.76$.   This is currently the highest redshift SMG known, suggesting that intense star formation was occurring in galaxies when the Universe was just over 1\,Gyr old.

We adopt cosmological parameters from the \textit{WMAP} fits in \citet{Spergel03}: $\Omega_\Lambda=0.73$, $\Omega_\mathrm{m}=0.27$, and $H_\mathrm{0}=71$\,km\,s$^{-1}$\,Mpc$^{-1}$.  At $z=4.76$ the angular scale is 6.5\,kpc arcsec$^{-1}$, and the Universe is 1.3\,Gyr old. All quoted magnitudes are on the AB system unless otherwise noted.

\section{Observations and Source Multiwavelength Properties}

%
%
\begin{figure*}
\epsfig{file=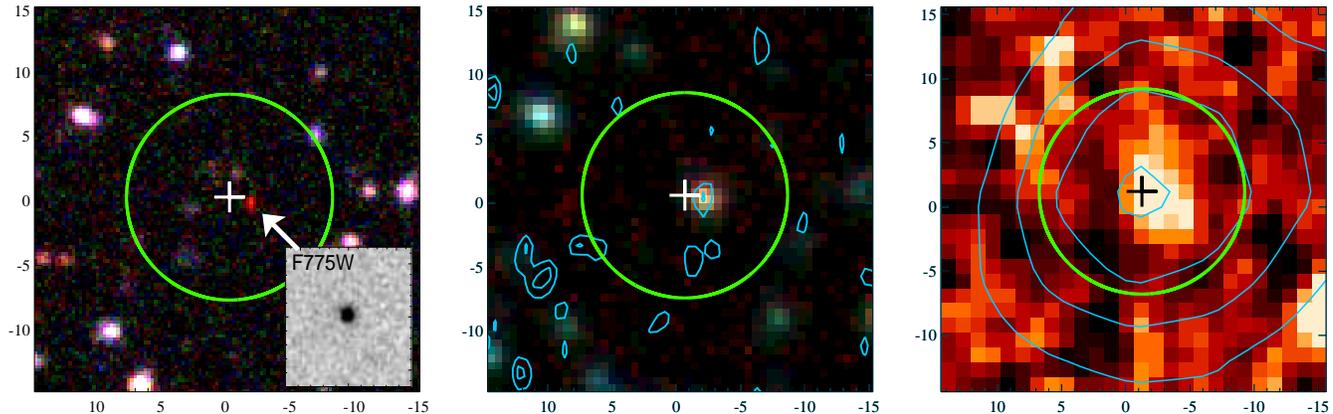,width=1.0\textwidth}
\caption{A selection of multiwavelength views of LESS\,J033229.4.    The three panels show: (left) a true-colour ground-based $BVR$ image of LESS\,J033229.4 from MUSYC, with a $2''\times2''$ inset illustrating the GOODS {\it HST} morphology in the F775W band (consistent with an unresolved point source); (centre) three-colour {\it Spitzer} 3.6, 4.5, 5.8\,$\mu$m imaging from GOODS/SIMPLE with the \citet{Miller08} VLA 1.4-GHz radio contours overlaid (starting at 2\,$\sigma$ and increasing in steps of 1\,$\sigma$); (right) the 24\,$\mu$m imaging from the {\it Spitzer}--FIDEL survey with the LABOCA 870$\mu$m emission from W09 overlaid as contours (starting at 2\,$\sigma$ and increasing in steps of 1\,$\sigma$).
The submillimetre position is indicated by a cross, and we also plot a circle representing the $8''$ search radius used to locate potential counterparts.  We have robustly identified the SMG with a radio and 24-$\mu$m counterpart $\sim1.5''$ away, which is coincident with a $z=4.76$ galaxy (see text).  Each panel is $15''\times15''$ in size and is orientated with North to the top and East to the left. 
}
\label{fig:stamp}
\end{figure*}

The full area of the ECDFS was mapped at 870\,$\mu$m with LABOCA on the 12-m Atacama Pathfinder EXperiment (APEX; \citealt{Gusten06}) telescope at Llano de Chajnantor in Chile as a joint ESO--Max Planck programme (the LESS collaboration).  A total of 310\,hrs of observing time (with 200\,hrs of that on-source) was obtained which yielded a map covering the central $30'\times30'$ of the field (0.25\,deg$^{2}$) to an average rms of $\sim1.2$\,mJy.  This is the largest contiguous deep submillimetre survey yet undertaken  (c.f.\ the $\sim0.25$\,deg$^{2}$ 2-mJy rms SCUBA Half Degree Extragalactic Survey -- SHADES; \citealt{Coppin06}). The description of the observations, reduction and analysis of the map is given in W09.  The robust catalogue derived from this map contains 121 SMGs at $>3.7$-$\sigma$ significance (chosen to provide an estimated false detection rate of $\leq 5$ sources), and the multiwavelength identification and characterisation of the SMG population in the ECDFS will be presented in subsequent papers.  Here we focus on the multiwavelength characteristics of a single source: LESS\,J033229.4$-$275619 (hereafter LESS\,J033229.4), detected at 870\,$\mu$m with a flux density of $6.3\pm1.2$\,mJy (S/N\,=\,5.1) at $03^\mathrm{h}32^\mathrm{m}29.41^\mathrm{s}$ $-27^{\circ}56'18.90''$ (J2000), with a 90 per cent positional uncertainty of $\sim 6''$ (W09).  W09 apply a flux-deboosting algorithm to correct the raw flux and noise values for all of the LESS sources in order to obtain the most accurate submm photometry (see e.g.~\citealt{Coppin05}).  In this paper we use the deboosted flux and noise of $5.0\pm1.4$\,mJy for LESS\,J033229.4.

We use a new radio catalogue (Biggs et al.\ in preparation) which comprises
emitters with peak flux densities in excess of 3\,$\sigma$, where
$\sigma$ is determined locally, extracted from the Very Large Array
(VLA) 1.4-GHz map of \citet{Miller08}. We search within $8''$ of the
submm position for potential radio counterparts (see e.g.~\citealt{Downes,Ivison02}) 
and identify a single candidate at a radial
distance of 1.5$''$ with a flux density of $24.0\pm6.3$\,$\mu$Jy
(S/N=3.8), assuming that it is a point source. The probability of this
radio emitter being aligned by chance with the SMG is less than 2 per
cent and we can assign this radio counterpart to the SMG with some
confidence. Radio images suffer from similar flux boosting effects as
the submm images and we thus adopt a deboosted 1.4-GHz flux density of
$18.8\pm6.3$\,$\mu$Jy from Biggs et al.\ (in preparation), following extensive
simulations of the kind described by \citet{Ibar09}.  The positional
accuracy of the radio data allows us to pinpoint LESS\,J033229.4 to
$\alpha_{\rm J2000}= 03^\mathrm{h}32^\mathrm {m}29.30^\mathrm{s}$
$\delta_{\rm J2000}=-27^{\circ}56^{'}19.40^{''}$ with an uncertainty
of $\Delta\alpha=0.2''$ and $\Delta\delta=0.3''$ (see
\citealt{Ivison07} and Fig.~\ref{fig:stamp}).  Using this precise
position for the robust radio counterpart of LESS\,J033229.4 we can
now use the impressive archival observations of the ECDFS to derive
its spectral energy distribution.

In the mid-to-far-infrared, a counterpart is detected at 3.6--$8\mu$m using the InfraRed Array Camera (IRAC; \citealt{Fazio04}) data from the Great Observatories Origins Deep Survey (GOODS\footnote{http://www.stsci.edu/science/goods/}; \citealt{goods}; \citealt{Giavalisco04}) imaging of the central part of the ECDFS (GOODS-South) and in the wider area coverage provided by the {\it Spitzer} IRAC/MUSYC Public Legacy Survey in the ECDFS (SIMPLE\footnote{http://data.spitzer.caltech.edu/popular/SIMPLE}; Damen et al.\ in preparation).  For the present analysis we use the 3-$''$ radius aperture photometry in the SIMPLE catalogue, as it provides close to optimal S/N for point sources, with the appropriate aperture corrections applied and corrected to total fluxes.  In the mid-infrared, LESS\,J033229.4 is detected by the Far-Infrared Deep Extragalactic Legacy survey (FIDEL; Dickinson et al.\ in preparation) in the Multiband Imaging Photometer for {\it Spitzer} (MIPS; \citealt{Rieke04}) 24\,$\mu$m catalogue at $03^\mathrm{h}32^\mathrm{m}29.30^\mathrm{s}$ $-27^{\circ}56^{'}18.78^{''}$, with a positional accuracy of 1--2$''$, given the FWHM and the source's S/N.  The MIPS source lies within $0.6''$ of the radio counterpart and hence the radio and MIPS emission originate from the same source, given the combined error circle.  We report a 24\,$\mu$m flux density of $S_{\rm 24\mu m}= 31.6\pm5.1 \mu$Jy (S/N\,=\,6.6), exploiting the deeper GOODS 24-$\mu$m imaging (Chary et al.\ in preparation). LESS\,J033229.4 is not detected at either 70 or 160\,$\mu$m in the deep FIDEL imaging and so we adopt 3-$\sigma$ upper limits at these wavelengths.

At shorter wavelengths, the region around LESS\,J033229.4 is covered by optical and near-infrared imaging from several ground and space-based public surveys: the Multiwavelength Survey by Yale--Chile (MUSYC\footnote{http://www.astro.yale.edu/MUSYC}; \citealt{Gawiser06}), the {\it Hubble Space Telescope (HST)} Advanced Camera for Surveys (ACS) Galaxy Evolution from Morphologies and SEDs survey (GEMS\footnote{http://www.mpia.de/GEMS/gems.htm}; \citealt{Rix04}), and GOODS (\citealt{Giavalisco04}; Giavalisco et al.\ in preparation). We list the fluxes or limits from these surveys in Table~\ref{tab:photom}. The source is undetected in all bands shortward of $\sim 600$\,nm, although it is significantly detected in the redder wavebands, $V_{606}, R, i_{775}, z_{850}$.  Unfortunately the source falls $\sim 1''$ outside the deep Very Large Telescope (VLT) Infrared Spectrometer And Array Camera (ISAAC) $JHK$ imaging of GOODS-South and so there are currently only weak constraints on its near-infrared fluxes from MUSYC \citep{Taylor08}. 

Finally, turning to the X-ray waveband: from their analysis of the 1-Ms {\it Chandra} image, \citet{Giacconi02} report the detection of an X-ray source at $03^\mathrm{h}32^\mathrm{m}29.44^\mathrm{s}$ $-27^{\circ}56''20.18''$ with a positional uncertainty of $2.1''$, with 0.5--2 and 2--10\,keV fluxes of $(1.6\pm 0.5) \times10^{-16}$ and $<7.6\times10^{-16}$\,erg\,s$^{-1}$\,cm$^{-2}$, respectively.  This  position is consistent with the GOODS optical counterpart $0.8''$ away.  However, this source (ID\,618) is only listed in their supplementary catalogue and is based on a SExtractor \citep{sextractor} analysis of the X-ray image, for which the software was not intended.   In contrast, no X-ray source is detected at this position in any of the analyses of the 250\,ks, 1-, and 2-Ms images of the CDFS,  by \citet{Lehmer05}, \citet{Alexander03a} or \citet{Luo08}, respectively. We thus extract 3-$\sigma$ upper limits from the deepest 2-Ms {\it Chandra} imaging of \citet{Luo08}, corresponding to rest-frame 3--11.5\,keV and 11.5---46\,keV luminosity limits of $<2.1\times10^{43}$ and $<1.5\times10^{44}$\,erg\,s$^{-1}$, respectively, which we use to constrain any contribution from an AGN to the bolometric luminosity (see \S~\ref{agn}).

\begin{figure}
\psfig{file=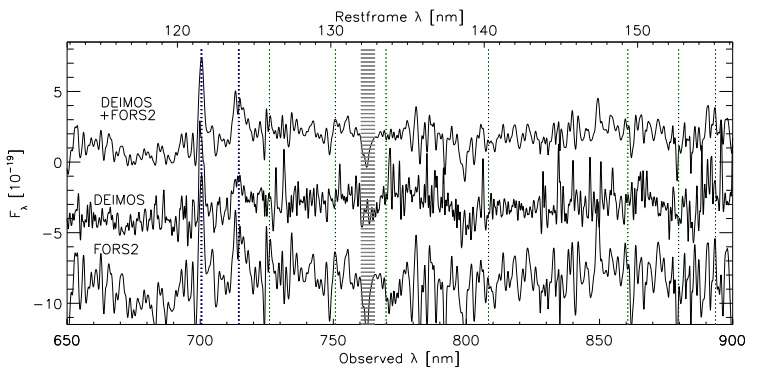,width=0.26\textwidth,angle=90}
\caption{The optical spectrum of LESS\,J033229.4 derived from combining our new deep DEIMOS spectrum with the archival FORS2 spectrum from \citet{Vanzella06}.  The upper spectrum is the combined dataset at the FORS2 resolution ($\sim 1.3$\,nm), while the lower spectra (arbitrarily offset in flux for clarity) show the individual DEIMOS spectrum convolved to $\sim 0.8$\,nm resolution and the FORS2 spectrum.  The combined spectrum displays two strong emission lines and an associated continuum break.  We identify the lines as Ly$\alpha$ and N{\sc v}, yielding a redshift of $z\sim 4.76$.  We mark these two emission features on the spectra and also indicate the expected wavelengths for other potential emission and absorption lines at the adopted redshift of $z=4.76$: Si{\sc ii}\,126.0, O{\sc i}\,130.3, C{\sc ii}\,133.6, Si{\sc iv}\,140.3, Fe{\sc iii}\,149.4, Si{\sc ii}\,152.7 and C{\sc iv}\,155.1, as well as atmospheric absorption at 760\,nm (shaded bar). The upper scale gives the rest-frame wavelength assuming a redshift of $z=4.76$, and the flux scale is nominally in erg\,cm$^{-2}$\,s$^{-1}$\,\AA$^{-1}$.}
\label{fig:spectrum}
\end{figure}

Table~\ref{tab:photom} summarises the fluxes or limits derived from the multiwavelength coverage of LESS\,J033229.4. From the optical photometry it is clear that the source is a $V$-band dropout.  Assuming this arises from the presence of the Lyman limit in the $V$-band, this suggests a redshift of $z\approx 5$. The morphology of the source in the redder {\it HST} imaging is very compact, with an observed FWHM of $\lsim 0.1''$ (or a half-light radius of $\lsim 0.3$\,kpc), indicating that the source is unresolved (see inset in Fig.~\ref{fig:stamp}).  Indeed, the source was photometrically and morphologically selected as a candidate $z=3.5$--5.2 AGN by  \citet{Fontanot07}. On this basis, LESS\,J033229.4 was observed with the FOcal Reducer and low dispersion Spectrograph (FORS2; \citealt{Appenzeller}) on the VLT by \citet{Vanzella06} (see also \citealt{Vanzella08}) as part of the ESO Large Programme 170.A-0788 (PI: Cesarsky). Their ID for this source is GDS\,J033229.29$-$275619.5 and they obtained a 4-hr exposure of it using the 300I grism, deriving a redshift of $z=4.76$ on the basis of the identification of strong Ly$\alpha$ emission, N{\sc v} and a continuum break (Fig.~\ref{fig:spectrum}).  The photometric properties of LESS\,J033229.4 are consistent with this claimed redshift, in particular the  discontinuity at $\sim 600$\,nm is well reproduced by the Lyman limit at $z=4.76$ (Fig.~\ref{fig:sed}).  This makes this source the highest redshift submillimetre-selected SMG yet discovered and indeed one of the highest redshift 24-$\mu$m-detected sources so far identified.

%
%
\begin{table}
\begin{minipage}{0.5\textwidth}
\scriptsize
\caption{The observed photometry for LESS\,J033229.4 from {\it Chandra}, {\it HST}, MUSYC, {\it Spitzer}, APEX and the VLA.  Note that we list the deboosted submm and radio fluxes, which is important in order to correctly model the SED and investigate the radio and submm properties of the SMG (see text).  We use the \textit{HST}-ACS photometry from \citet{Stark07}, which has already been corrected for the Ly$\alpha$ emission measured from the FORS2 rest-frame UV spectrum.  We also give the $3.6\,\mu$m flux density, corrected down by 30 per cent to reflect the typical equivalent width of H$\alpha$ in $z\sim2$ SMGs (\citealt{Swinbank04}; \citealt{Takata06}).  Where the source is undetected we quote a 3-$\sigma$ upper limit to the flux density. Note: 1$\mu$Jy is the equivalent of 23.9 AB magnitudes.}
\label{tab:photom}
\hspace{-0.2in}
\begin{tabular}{lll}
\hline
\multicolumn{1}{c}{Filter} & \multicolumn{1}{c}{Flux Density} & \multicolumn{1}{c}{Reference} \\ 
\multicolumn{1}{c}{/Wavelength} & \multicolumn{1}{c}{($\mu$\,Jy)} & \multicolumn{1}{c}{} \\ 
\hline
0.5--8\,keV & $<3.7\times10^{-16}$\,erg\,s$^{-1}$\,cm$^{-2}$ & \citet{Luo08} \\ 
0.5--2\,keV & $<8.5\times10^{-17}$\,erg\,s$^{-1}$\,cm$^{-2}$ & \citet{Luo08} \\
2--8\,keV   & $<6.3\times10^{-16}$\,erg\,s$^{-1}$\,cm$^{-2}$ & \citet{Luo08} \\
$U$ (352\,nm)     & $<0.087$ & \citet{Gawiser06} \\
$B_{435}$ (433\,nm)     & $<0.013$ & \citet{Fontanot07} \\
$B$ (461\,nm)     & $<0.039$ & \citet{Gawiser06} \\ 
$V$ (538\,nm)     & $<0.059$ & \citet{Gawiser06} \\
$V_{606}$ (597\,nm)     & $0.072\pm0.006$ & \citet{Stark07} \\
$R$ (652\,nm)     & $0.215\pm0.020$ & \citet{Gawiser06} \\
$i_{775}$ (771\,nm)     & $0.350\pm0.013$ & \citet{Stark07} \\
$z_{850}$ (905\,nm)     & $0.405\pm0.015$ & \citet{Stark07} \\
$J$ (1.2$\mu$m)     & $<1.65$ & \citet{Taylor08} \\
$K$ (2.1$\mu$m)     & $<3.20$ & \citet{Taylor08} \\
3.6\,$\mu$m & $2.86\pm0.06$ & Damen et al.\ in preparation \\
4.5\,$\mu$m & $4.04\pm0.08$ & Damen et al.\ in preparation \\
5.8\,$\mu$m & $6.3\pm0.4$ & Damen et al.\ in preparation \\
8.0\,$\mu$m & $9.2\pm0.4$ & Damen et al.\ in preparation \\
24\,$\mu$m  & $32\pm5$ & Chary et al.\ in preparation \\
70\,$\mu$m  & $<2500$ & Dickinson et al.\ in preparation \\
160\,$\mu$m  & $<33000$ & Dickinson et al.\ in preparation \\
870\,$\mu$m  & $5000\pm1400$ & W09 \\
20\,cm       & $18.8\pm6.3$ & Biggs et al.\ in preparation \\
\hline
\end{tabular}
\end{minipage}
\end{table}
\normalsize

%
%

We obtained additional spectroscopic observations of LESS\,J033229.4 using the DEep Imaging Multi-Object Spectrograph (DEIMOS; \citealt{Faber03}) on Keck on the nights of 2007 October 10 and 11.  The total exposure time was 6.5\,hrs in good conditions using the 600 lines\,mm$^{-1}$ grating and the resulting spectrum (Fig.~\ref{fig:spectrum}) shows continuum emission and the two strong emission lines seen in the FORS2 spectrum.  We retrieved the reduced FORS2 spectrum of \citet{Vanzella06} from the ESO archive and then convolved the DEIMOS spectrum to the $\sim 1.3$\,nm resolution of the FORS2 spectrum before averaging the two spectra.  The combined spectrum has an effective 10-m equivalent integration time of $\sim 9$\,hrs and is shown in Fig.~\ref{fig:spectrum}.  The combined spectrum displays a narrow Ly$\alpha$ emission line at 700.5\,nm and a broad N{\sc v} emission line at 714.5\,nm with a width of $\sim5$\,nm or $\sim2000$\,km\,s$^{-1}$ in the rest-frame and an integrated flux density comparable to Ly$\alpha$ ($\sim 1.4\times 10^{-17}$\,erg\,s$^{-1}$\,cm$^{-2}$).  To search for weak broad lines we have convolved the spectrum to the FWHM of the N{\sc v} line and find an emission feature corresponding to C{\sc iv}\,155.0 at very low significance, likely due to the strong telluric emission at these wavelengths.  The spectral properties of the source, in particular the relatively strong and moderately broad N{\sc v}, indicates the presence of an AGN, though the weak X-ray and 24\,$\mu$m flux densities relative to the optical and far-infrared bands suggest a relatively weak AGN contribution to the bolometric luminosity.  

High-redshift AGN are often associated with rest-frame far-infrared emission (e.g.\ \citealt{Ivison08}), and as a final test of our identification of the SMG with this source, we can also ask what is the likelihood of finding such a high-redshift source close to the submillimetre position by chance. Using the AGN luminosity function from \citet{Fontanot07}, we calculate how many $z>4$ AGN counterparts we would have expected to find at random given their volume density and the offset of the submillimetre source to the AGN, $\theta=1.5''$.  LESS\,J033229.4 has a rest-frame 145-nm luminosity of M$_{145}=-21.5$, calculated following \citet{Fan00} and adopting $f_{\nu} \propto \nu^{-\alpha}$ with $\alpha=0.5$.  The volume density of AGN brighter than LESS\,J033229.4 is $\Phi$(M$_{145}<-21.5) \sim 2.6\,\times 10^{-6}$\,Mpc$^{-3}$, equivalent to a surface density $N\sim34$\,deg$^{-2}$ for the integrated density at $z>4$.  The Poisson random probability of finding a $z>4$ AGN at least as bright as ours is $P=1-\mathrm{exp}[-\pi \theta^{2} N]\sim 10^{-5}$. Therefore it is extremely unlikely that our SMG is located within $1.5''$ of a high-redshift source by random chance, confirming the proposed association of the submillimetre emission with the $z=4.76$ source.

\section{Results \& Analysis}

\subsection{Spectral Energy Distribution}\label{sed}

%
%
\begin{figure*}
\psfig{file=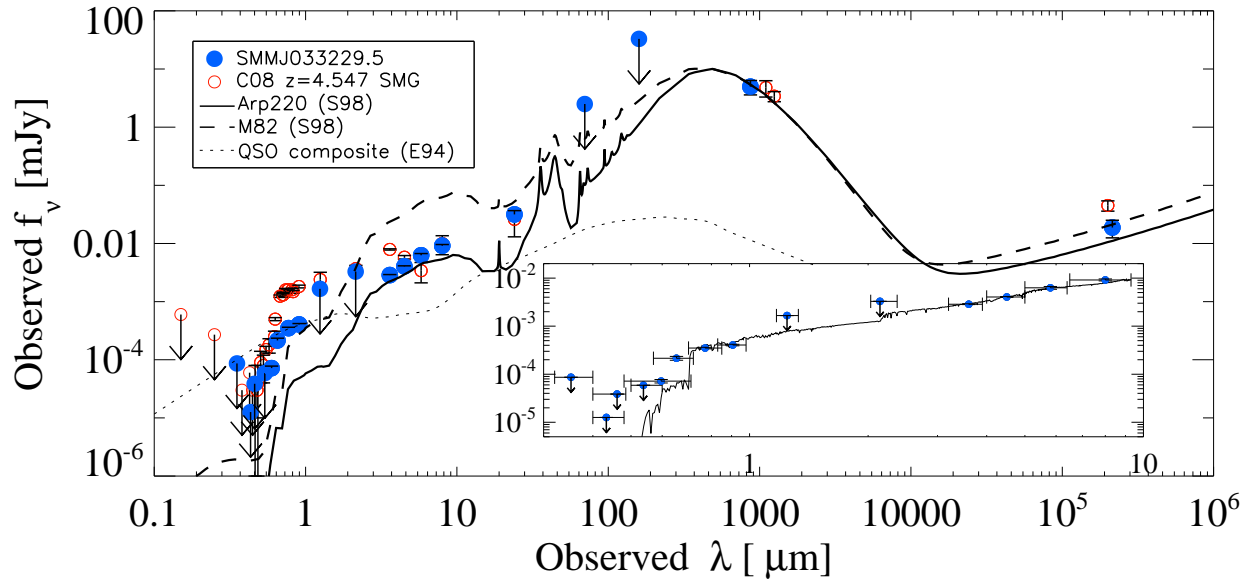,width=1.0\textwidth}
\caption{The SED of LESS\,J033229.4 showing the dominant and luminous rest-frame far-infrared emission indicating that the source is an ultraluminous infrared galaxy. For comparison we have overplotted the SED of Arp\,220 and M\,82 from \citet{Silva98} redshifted to $z=4.76$ and scaled to match the $S_\mathrm{870\,\mu m}$ flux density of LESS\,J033229.4 and also a composite QSO SED from \citet{Elvis} redshifted to $z=4.76$ scaled to match the rest-frame UV flux density. The good agreement with the starburst templates in the far-infrared and radio suggest that the bulk of the  bolometric emission in this system arises from star formation.  We also plot the SED of the $z=4.5$ SMG SMM\,J100054.5+023436 from Capak et al.\ (2008; C08) to show the similar far-infrared properties, but the much more intense rest-frame UV emission from that system. The inset (in the same units as the larger plot) shows the best-fitting stellar SED from {\sc HyperZ} to the optical--$8\,\mu$m photometry, a 40-Myr old constant star formation model with moderate reddening, $A_V\sim 1.5$, which provides an acceptable fit of the rest-frame UV--optical emission in LESS\,J033229.4.  However, we caution that the current SED can be equally well-fit by a power-law, as expected from a reddened AGN, leaving open the possibility that a significant fraction of the rest-frame near-infrared emission arises from the AGN.}
\label{fig:sed}
\end{figure*}

The bright submillimetre flux density of LESS\,J033229.4 suggests
that a massive burst of star formation is underway in this system at $z=4.76$.   
To test if the observed radio and far-infrared emission is  consistent with
being produced by star formation, we calculate the radio-to-submm
spectral index for LESS\,J033229.4,  
deriving $\alpha^{350}_{1.4}=1.01\pm0.11$.   This spectral
index is in good agreement with the
mean spectral index at $z=4.76$ predicted from 17
local star-forming galaxies by \citet{Carilli00}: 
$\alpha^{350}_{1.4}=1.04\pm 0.13$,
as well as with local ULIRGs such as Arp\,220, $\alpha^{350}_{1.4} \sim 1.09$. 
This confirms that the far-infrared--radio emission from LESS\,J033229.4 
is consistent with massive star formation.

We therefore plot the multiwavelength spectral energy distribution (SED) of LESS\,J033229.4 in
Fig.~\ref{fig:sed} and compare this to the SEDs of  Arp\,220 and M\,82,
normalised to match the observed $S_\mathrm{870\,\mu m}$ of
LESS\,J033229.4 -- corresponding to rest-frame 150-$\mu$m emission,
near the peak of the dust emission. This shows that while the Arp\,220 
and M\,82 SEDs both fit the radio--far-infrared emission from 
LESS\,J033229.4, M\,82 is significantly brighter in the mid-infrared,
with only Arp\,220 providing an adequate description of the 
SED across the rest-frame optical--to--radio.  This confirms that this $z=4.76$ SMG has an SED
consistent with local star-formation-dominated ULIRGs, although we
note that the rest-frame UV and 24\,$\mu$m emission exceed that seen
for Arp\,220.  Therefore we have also overplotted a redshifted
composite QSO SED from the atlas of local, optically bright
quasars compiled by \citet{Elvis}, scaled to match our observed rest-frame UV
photometry, in order to assess the likely bolometric contribution of an unobscured AGN to our observed SED (see Fig.~\ref{fig:sed}). 
It is apparent from this simple exercise that any contribution from an unobscured AGN 
to the infrared luminosity of LESS\,J033229.4 is likely to be negligible, and we
investigate this in detail in \S\,\ref{agn}.

From the scaled Arp\,220 template, we calculate
$L_\mathrm{FIR}=L_\mathrm{(8-1000\,\mu
m)}=6.1\times10^{12}$\,L$_{\odot}$, a factor $\sim5 \times$ the
bolometric luminosity of Arp\,220 (\citealt{Soifer84};
\citealt{Sanders88,Sanders03}).  We note that adopting the scaled M\,82 
template results in an $L_\mathrm{FIR}$ a factor of $\sim1.5\times$ higher. 
Owing to the negative K correction in the submillimetre
waveband, the bolometric luminosity of LESS\,J033229.4 is close to the median luminosity of the $z\sim 2$ population in
\citet{Chapman05} (see also \citealt{Kovacs06}; \citealt{Coppin08a}),
even though its luminosity distance is $2.5\times$ larger. 

Based on these SED arguments, which 
indicate that much of the far-infrared emission arises from star formation,
we derive a SFR of $\sim
1000$\,M$_\odot$\,yr$^{-1}$ for LESS\,J033229.4, following
\citet{Kennicutt98} who assumes a burst lifetime
of $\sim 100$\,Myr and a \citet{Salpeter55} initial mass function
(IMF).  Next, we estimate a dust mass of $M_\mathrm{d}\sim5\times10^{8}\,M_{\odot}$
using the 870\,$\mu$m flux density and the usual relation
$S_{\nu_\mathrm{obs}}=B_{\nu'}(T)\kappa_{\nu'}\,M_\mathrm{d}\,(1+z)/D_{\mathrm{L}}^{2}$
(e.g.~\citealt{HDR97}), where $D_{\mathrm{L}}$ is the cosmological
luminosity distance and $B_{\nu'}$ is the Planck function evaluated at
the emitted frequency, $\nu'=\nu_\mathrm{obs}\,(1+z)$.  We have
extrapolated from an average wavelength-dependent mass-absorption
coefficient of $\kappa_{125_\mathrm{\mu m}}=2.64\,\mathrm{m}^{2}\,\mathrm{kg}^{-1}$ assuming $\beta=1.5$
\citep{Dunne03}.  Based on the similar far-infrared luminosity and
spectral properties of LESS\,J033229.4 to $z\sim 2$ SMGs we adopt the
average gas/dust ratio for this population ($\sim 60$;
\citealt{Greve05}; \citealt{Tacconi06}; \citealt{Kovacs06};
\citealt{Coppin08a}), and so predict a gas mass for LESS\,J033229.4
of $\sim 3\times10^{10}$\,M$_\odot$. 
Comparing this mass to the SFR, we expect  that LESS\,J033229.4
would consume its gas at its present SFR in $\sim30$--40\,Myr.  This
predicted gas mass is similar to that determined for other $z>4$ SMGs
(\citealt{Daddi08}; \citealt{Schinnerer08}).

To place limits on the potential stellar mass of the galaxy, 
we now investigate if the rest-frame UV--optical emission can be fit by a stellar population using the {\sc
HyperZ} package \citep{Bolzonella00}. We stress that this is a necessary, but not a sufficient, requirement 
to show that this emission actually arises from stars, rather than the AGN.
We assume a solar metallicity stellar population model from
\citet{BruzChar93}, a \citet{Calzetti00} starburst attenuation law, and
either a constant or single-burst star formation history. 
Under these assumptions, {\sc HyperZ}
shows that the observed SED can be fit by a stellar
population at $z=4.76$, although statistically a reddened power-law provides
as good a fit.  The
best-fit SED is a 40-Myr old constant star formation model with
moderate reddening, $A_{V}\sim1.5$ (see inset
Fig.~\ref{fig:sed}). However, the absence of useful limits on the
galaxy's SED in the observed $JHK$ bands, sampling the critical region
around the Balmer and 4000\AA\ breaks, means that there is
significant uncertainty in these parameters, with statistically acceptable
solutions with ages up to $\gsim 500$\,Myr and lower reddening,
$A_V\lsim 0.5$.  Even this range of fits should be used with caution
as the potentially dominant contribution from the AGN, as well as
the complex mix of dust  and stars, in this SMG will mean
that the luminosity-weighted ages and reddening will vary strongly
with wavelength as different regions within the galaxy are
probed. Hence, in the following we simply assume a canonical duration of
the SMG-phase of $\sim 100$\,Myr  \citep{Swinbank06}, and focus on the rest-frame near-infrared
luminosity which should suffer the least AGN contamination and provide the most robust estimate of the stellar mass of the galaxy.

From a simple interpolation of the SED we derive a rest-frame $K$-band luminosity of M$_K\sim -24.0$ or L$_K\sim 5\times 10^{11}$\,L$_\odot$. 
To estimate a limit on the stellar mass from this we use the parameterisation for L$_K/$\,M$_\odot$ for a constant star formation model from \citet{Borys05}.  For star formation which has been occuring for 100\,Myr this gives L$_K/$\,M$_\star \sim 10$ (40- and 300-Myr durations correspond to L$_K/$\,M$_{\odot} \sim 50$ and 7), implying a typical stellar mass of M$\star\lsim 5\times 10^{10}$\,M$_\odot$.  This has an uncertainty of at least a factor of 3--$5\times$, even before considering the potential contribution from an AGN to the rest-frame near-infrared emission.
Nevertheless, we note that this limit on the stellar mass is comparable to the predicted gas mass of the system and that for ages of $\sim 100$\,Myr, the stellar mass produced in the burst implies a SFR of $\sim 500$\,M$_\odot$\,yr$^{-1}$, similar to that derived from the far-infrared.  
In contrast, \citet{Stark07} estimate a stellar mass for LESS\,J033229.4 of M$_{\star}\sim 1.3\times10^{11}$\,M$_\odot$, assuming the mass-to-light ratio appropriate for a $\sim 1$-Gyr old stellar population, which is unlikely to be valid for a system with a current SFR of $\sim 1000$\,M$_\odot$\,yr$^{-1}$.

\subsection{Constraints on AGN energetics}\label{agn}

The rest-frame UV compact morphology (FWHM of $\lsim 0.1''$; or a half-light radius of $\lsim 0.3$\,kpc) 
and spectroscopy of LESS\,J033229.4 indicate that it hosts an AGN.  For comparison, about $\sim10$ per cent of $z\sim2$ SMGs 
are classed as `compact', with half-light radii of $<0.6$\,kpc \citep{Swinbank09}.  
Similar spectral signatures of AGN emission are also seen in the rest-frame UV spectra of around 25 per cent of
the radio-detected SMG population at $z\sim 2$--3 \citep{Chapman05},
suggesting a mix of star formation and AGN in the galaxy. In particular,
LESS\,J033229.4's spectral properties are very similar to the first
spectroscopically identified SMG: the type-2 or narrow-line QSO
SMM\,J02399$-$0136 at $z=2.803$ \citep{Ivison98a}. SMM\,J02399$-$0136
also shows strong Ly$\alpha$ and a comparably bright and broad N{\sc v}
line, with a width of 1800\,km\,s$^{-1}$. Although the emission lines in SMM\,J02399$-$0136
are proportionally stronger relative to the continuum and the source
is intrinsically more luminous than LESS\,J033229.4.  

We can use the strength of the N{\sc v} emission and other 
SED constraints to estimate  the potential luminosity of the AGN
component in LESS\,J033229.4 and compare these to our X-ray
limits, where we are confident that any AGN emission will dominate.
Using the observed 2--8\,keV limit (rest-frame
11.5--46\,keV) and assuming $\Gamma=2.0$
we derive a conservative limit on the rest-frame
2--10\,keV luminosity of
$<1.7\times10^{44}$\,erg\,s$^{-1}$. 
This will be almost unaffected by the presence of
absorption assuming absorbing column densities of
$N_\mathrm{H}\lsim10^{25}$\,cm$^{-2}$, i.e. not heavily Compton
thick.  A more sensitive limit comes from 
the observed 0.5--2\,keV limit which yields a rest-frame
2--10\,keV luminosity limit of
$<2.5\times10^{43}$\,erg\,s$^{-1}$. However, this is also more sensitive
to the assumed absorption in the system and so we will use our more
conservative limit in the following.

Starting from 
the measured N{\sc v} emission-line luminosity of
$\sim3\times10^{42}$\,erg\,s$^{-1}$, we 
can use the observed N{\sc v}--X-ray luminosity ratio found for
SMM\,J02399$-$0136 (\citealt{Ivison98a}; \citealt{Bautz00}) to
predict the
rest-frame 2--10\,keV luminosity of LESS\,J033229.4 of 
$\sim1\times10^{44}$\,erg\,s$^{-1}$.  
We can also attempt to estimate the luminosity of the AGN using the 24\,$\mu$m flux
density under the extreme assumption that the 24\,$\mu$m emission 
is entirely due to the AGN.
Using the  derived rest-frame 6\,$\mu$m luminosity of
$1.2\times10^{45}$\,erg\,s$^{-1}$ 
we estimate a rest-frame 2--10\,keV luminosity of
$\approx3\times10^{44}$\,erg\,s$^{-1}$ based on \citet{Alexander08b}. 
Similarly, assuming that the UV continuum luminosity of LESS\,J033229.4 
is dominated by emission from the AGN, we can predict the AGN
luminosity. We take the observed rest-frame 
145-nm absolute magnitude ($M_{145}=-21.5$) and convert this to
rest-frame 250-nm  assuming $\alpha=-0.44$
($S_\nu \propto \nu^{\alpha}$; \citealt{VandenBerk01}).
We then use the
$\alpha_{\rm ox}$ relationship of \citet{Steffen06} and
convert between different X-ray bands assuming
$\Gamma=2.0$, to derive a rest-frame
2--10\,keV luminosity of $5\times10^{43}$\,erg\,s$^{-1}$.

These three estimates are in the range 0.5--$3\times 10^{44}$\,erg\,s$^{-1}$ and so are consistent with
our observed X-ray limit, given the significant uncertainties in the calculations. 

\section{Discussion}

Our estimates of the stellar and gas masses of LESS\,J033229.4 in \S\,\ref{sed} are both crude and highly uncertain; nevertheless, both masses are comparable to those derived for the better-constrained and similarly bolometrically luminous $z\sim 2$ SMG population.  The implied baryonic mass of the galaxy is thus likely to be $\sim 10^{11}$\,M$_\odot$, suggesting that the ultraluminous starburst is occuring in a massive galaxy. Indeed, similar constraints on the host galaxy masses have been derived for the other examples of $z>4$ SMGs (e.g.\ \citealt{Daddi08}; \citealt{Schinnerer08}).

Our three predictions of the X-ray luminosity of the AGN in \S\,\ref{agn} 
suggest the luminosity of the AGN in LESS\,J033229.4
is $\sim 10^{44}$\,erg\,s$^{-1}$ and that the AGN may dominate
the emission in the X-ray and UV, while making a modest contribution to the
mid-infrared luminosity.  Using the X-ray--to--infrared luminosity ratio for
local AGN (e.g.\ \citealt{Alexander05}) we expect that the AGN contributes $\lsim20$ per cent of the
bolometric luminosity and hence the bulk of the  
bolometric luminosity of LESS\,J033229.4 arises from star formation.

The discovery of LESS\,J033229.4 at $z=4.76$ and three $z=4$--4.5 SMGs from \citet{Capak08} and \citet{Daddi08} indicates that galaxies with SFRs of $\sim 1000$\,M$_\odot$\,yr$^{-1}$ are present at high redshifts. Using these four $z>4$ SMGs, and the areas of the surveys they were derived from, we can place a lower limit on the surface density of SMGs at these redshifts of $\gsim 6$\,deg$^{-2}$, or a volume density of $\gsim 1.5\times 10^{-7}$\,Mpc$^{-3}$ conservatively assuming the surveys are uniformly sensitive across  $z=4$--8.  Note that if we include the likely $z>4$ SMG from \citet{Wang07} (see also \citealt{Wang09}), then these limits increase by $\sim 15$ per cent.

To test whether these high-redshift SMGs can evolve into luminous red galaxies at $z\sim 3$, we compute the expected fading of LESS\,J033229.4 from $z\sim 4.5$ to $z\sim 3$.  Qualitatively, the inferred SFRs of the four $z>4$ SMGs ($\sim10^3$\,M$_\odot$\,yr$^{-1}$), with estimated duration of order 100\,Myr, are sufficient to form $>10^{11}$\,M$_\odot$ of stellar mass needed to match the properties of luminous red galaxies seen at $z\sim3$, and there is enough time between $z\sim4$ and $z\sim3$ to age their stellar populations sufficiently.  To quantitatively confirm this we adopt an approach that, while uncertain, is insensitive to the AGN contribution to the rest-frame $K$-band luminosity of  LESS\,J033229.4.
We assume a constant star formation rate at the observed level of 1000\,M$_{\odot}$\,yr$^{-1}$ for 100\,Myr and that the starburst is observed half way through, then we estimate fading after $\sim 1$--1.5\,Gyr of $\Delta K\sim1.1$--1.3 in the rest-frame $K$-band resulting in a typical SMG-descendent at $z\sim3$ with $M_K\sim -24.7\pm 0.1$ and observed colours of $(J-K)\sim 0.6$--0.9, assuming \citet{BruzChar93} stellar models. Using the \citet{Maraston98} stellar population models these predictions become $\Delta K\sim 0.7$--1.7, $M_K\sim -24.7\pm 0.5$ and $(J-K)\sim 1.0$--1.3. These predicted luminosities and colours are in reasonable agreement with those of the reddest luminous galaxies at $z\sim 2.5$--3.5, which have $M_K\lsim -23.5\pm 0.5$ and $(J-K)\geq 1.3$ (e.g.\ \citealt{Marchesini08}).  

We can also ask whether the space densities of these two populations are consistent with this evolutionary cycle. Our lower limit on the space density of SMGs at $z\sim 4$--8 is 
$\gsim 1.5\times 10^{-7}$\,Mpc$^{-3}$, so correcting for the duty cycle assuming a 100-Myr lifetime ($\sim 9\times$) this suggests
that there will be  $\gsim 10^{-6}$\,Mpc$^{-3}$ descendents at $z\sim 2.5$--3.5.  This compares 
to the measured space density of luminous galaxies at these redshifts of $\sim 5\times 10^{-5}$\,Mpc$^{-3}$
of which $\sim 20$--30 per cent have red colours  (\citealt{Marchesini07}; \citealt{Marchesini08}), or $\sim 10^{-5}$\,Mpc$^{-3}$.
This suggests that we have so far only detected $\sim 10$ per cent of the SMG progenitors at $z>4$ of the $z\sim 3$ evolved luminous galaxies if all of the latter have a previous ultraluminous phase (c.f.~\citealt{Stark09}), or alternatively that there
should be a surface density of $\sim 100$\,deg$^{-2}$ SMGs at $z>4$, corresponding to $\sim 10$ per cent of the $\gsim 5$\,mJy SMG
population. 

\begin{figure}
\epsfig{file=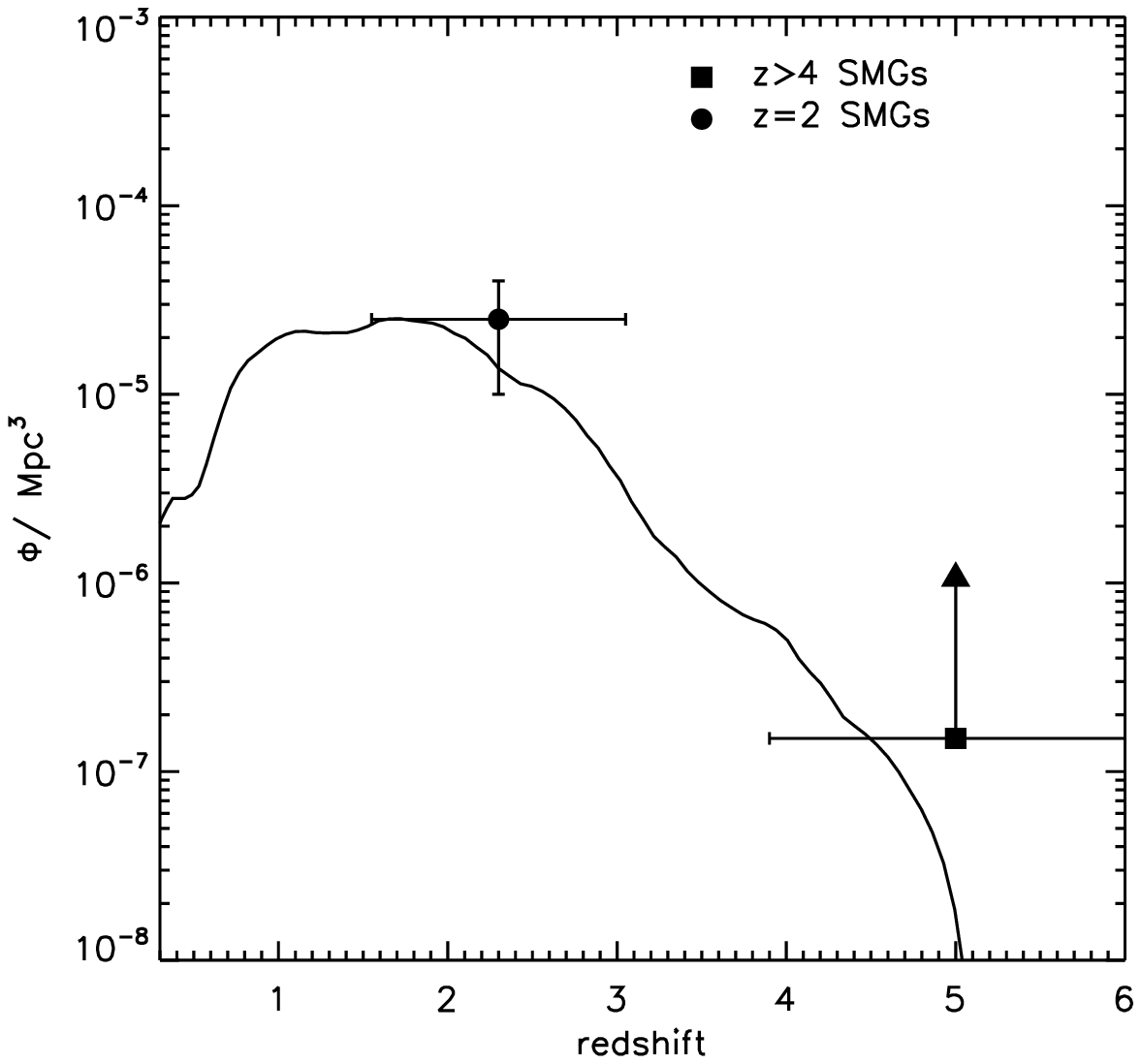,width=0.4\textwidth,angle=90}
\caption{A comparison of the volume density of radio-detected SMGs with 
$S_\mathrm{1.4\,GHz}>30\,\mu$Jy and  $S_\mathrm{850\,\mu m}>5$\,mJy from the GALFORM model of \citet{Baugh05}
to our limit on the volume density of $z>4$ SMGs and  the measured density of $z\sim 2$ SMGs from \citet{Chapman05}. 
This demonstrates that these detailed theoretical models are just consistent with these current observations, and so the discovery of more examples of $z>4$ SMGs within the ECDFS would result in a disagreement with the model predictions.
}
\label{fig:genzel}
\end{figure}

More detailed theoretical galaxy formation models have achieved a rough agreement with the number counts and the median redshift of the SMG population (e.g.\ \citealt{Baugh05}; 
\citealt{Swinbank08}).  The \citet{Baugh05} semi-analytical galaxy formation model implements a top-heavy IMF in bursts in order to reproduce the submillimetre number counts and has been shown by \citet{Swinbank08} to provide a reasonable description of the far-infrared properties of submillimetre galaxies at $z\sim 2$. 
We now investigate if the proportion of spectroscopically-confirmed high-redshift SMGs can provide a more stringent test of this model.  In Fig.~\ref{fig:genzel} we  compare the volume densities of spectroscopically-confirmed SMGs to the  model predictions (\citealt{Baugh05}).  \citet{Baugh05} predict surface and volume densities of $\gsim 7$\,deg$^{-2}$ and $\gsim 2\times 10^{-7}$\,Mpc$^{-3}$ (for $z\sim 4$--8), respectively, for SMGs with $S_\mathrm{850\,\mu m}>5$\,mJy (see \citealt{Swinbank08}).  Imposing a radio-detection constraint ($\gsim30$\,$\mu$Jy), which can be directly compared against the SMG samples with spectroscopic confirmation, reduces these predictions by a factor of $\sim2$ (Fig.~\ref{fig:genzel}).  Both of these estimates are consistent with our current limits on the volume densities of SMGs at $z>4$ (as well as the measured volume density at $z\sim 2$ from \citealt{Chapman05}).  However, we note that the follow-up of the LABOCA survey of the ECDFS has only just begun and so it is possible that more
$z>4$ SMGs will be uncovered.  Hence,  we suggest that it is probable that the current semi-analytic models underpredict the numbers of very high redshift ULIRGs.   In particular, if an example of a $z>5$ SMG is found in existing submillimetre surveys then this will place strong constraints on these models.

%
%

\section{Conclusions}

LESS\,J033229.4 is currently the highest redshift submillimetre-selected galaxy known.  At $z=4.76$, it is seen just over 1\,Gyr after the Big Bang, yet it has similar properties to the SMG population seen at the peak of their activity at $z\sim 2$ when the Universe was nearly $3\times$ older.  The SED and rest-frame UV spectrum indicate that LESS\,J033229.4 is a composite AGN/starburst galaxy caught in a phase where star formation (SFR\,$\sim 1000$\,M$_\odot$\,yr$^{-1}$) is dominating the bolometric emission.  

At $z=4.76$, 870\,$\mu$m samples near the SED peak, providing a good constraint on the far-infrared luminosity of SMG of $6\times10^{12}$\,L$_\odot$. But in order to accurately determine the SED peak shape and bulk dust temperature of LESS\,J033229.4 one requires photometry on the Wien side of the peak of the SED from  observations of the ECDFS at 250, 350 or 500\,$\mu$m with the {\it Herschel Space Observatory} or the Balloon-borne Large Aperture Submillimetre Telescope (BLAST; \citealt{Pascale08}).  Our preferred Arp\,220-like SED fit predicts 250, 350 and 500\,$\mu$m flux densities of $\sim 5$, 9, and 10\,mJy at these wavelengths, which are likely to prove to be challenging for either {\it Herschel} or BLAST due to confusion. Although a deep 350\,$\mu$m (rest-frame 60$\mu$m) or 450$\mu$m flux density measurement would be feasible with future APEX measurements using the SABOCA submillimetre camera or the James Clerk Maxwell Telescope's Submillimetre Common-User Bolometer Array 2 (SCUBA-2), respectively.

Based on our estimated dust mass of $5\times10^{8}$\,M$_\odot$ and adopting a $z\sim2$ SMG average gas/dust ratio, we predict a gas mass of $3\times10^{10}$\,M$_\odot$, similar to that measured for other $z>4$ SMGs (\citealt{Daddi08}; \citealt{Schinnerer08}).  At $z=4.76$ $^{12}$CO(5--4) emission should be detectable using current interferometric facilities.  The confirmation of a massive reservoir of gas in LESS\,J033229.4 would enable the study of the gas and dynamics of galaxy formation in detail when the Universe was only 1\,Gyr old, and so better place LESS\,J033229.4 in context with SMGs and red luminous galaxies at $z\sim3$.

As with the original identification of SMGs \citep{Ivison98a}, it is likely that selection effects may be strongly influencing our view of the properties of the very high-redshift SMG population.  We note that if LESS\,J033229.4 had not been identified as a compact, potential high-redshift, AGN then it would not have been spectroscopically observed and if it had not shown strong line emission (at least in part from the AGN) then a redshift would never have been measured. Similarly, the  morphologically complex optical counterpart of the \citet{Capak08} SMG was first identified as a weak radio source associated with a potentially $z>4$ $V$-band dropout object, and the redshifts for the two $z=4.05$ \citet{Daddi08} SMGs were obtained from the serendipitous detection of redshifted $^{12}$CO(4--3) emission in observations of an unrelated nearby $z=1.5$ BzK galaxy.  A complete view of the properties of the $z>4$ SMG population will require a concerted identification programme and multiwavelength follow-up, but not withstanding this, at present we can conclude that there appears to be an similarly diverse population of intense star-bursting galaxies at $z\gsim 4$ as there is at $z\sim 2$. 

Assuming that red luminous galaxies at $z\sim3$ are all descendants of $z=4$--8 SMGs we estimate the space density of the
SMG progenitor population would be of the order of $\sim 10^{-5}$\,Mpc$^{-3}$ at $z\sim 4$--8, assuming a 100-Myr lifetime
for the ULIRG-phase, corresponding to a surface density of $\sim 50$--100\,deg$^{-2}$.  Current observational limits on the $z>4$ SMG population are thus consistent with them being the progenitors of $>10$\% of the passive massive galaxy population at $z\sim 3$.  These
limits are in agreement with the semi-analytical model predictions for radio-detected SMGs at these redshifts.  However, we note that 
if further follow-up shows that the surface density of SMGs at $z\sim 4$--8 is closer to the $\sim 100$\,deg$^{-2}$
necessary for them to form all of the red luminous galaxy population at $z\sim 3$, then the models will be underpredicting the number of SMGs at $z>4$ by about a order of magnitude.  Alternatively, many $z\sim3$ red luminous galaxies may form without a superluminous phase.  Either way, statistically significant samples of SMGs need to be studied and followed up with spectroscopy in fields with extensive multiwavelength coverage before we will be able to determine true predominance and role of SMGs at $z>4$.  The panoramic and deep Cosmology Legacy Surveys planned with SCUBA-2 will obtain the tens of thousands of SMGs needed to begin the search in earnest for SMGs at even higher redshifts, $z\gg 4$--8.

\section{Acknowledgments}\label{ack}

We would like to thank an anonymous referee for helpful comments and suggestions on the paper.  We thank John Helly for help with extracting information from the Millenium and GALFORM databases.  KEKC acknowledges support from a Science and Technology Facilities Council (STFC) fellowship.  IRS, DMA and JSD acknowledge support from the Royal Society.  JEG and JLW acknowledge support from the STFC, and AMS acknowledges support from a Lockyer fellowship.  The work of DS was carried out at Jet Propulsion Laboratory (JPL), California Institute of Technology (Caltech), under a contract with National Aeronautics and Space Administration (NASA).  
WNB acknowledges support from the {\it Chandra} X-ray Observatory's grant SP8-9003A.

Observations have been carried out using APEX and the VLT under Program IDs: 080.A-3023, 079.F-9500, 170.A-0788, 074.A-0709, and 275.A-5060.  APEX is operated by the Max-Planck-Institut f\"{u}r Radioastronomie, the European Southern Observatory, and the Onsala Space Observatory.  We would like to thank the staff for their aid in carrying out the observations.  Some of the data presented herein were obtained at the W.\,M.\ Keck Observatory, which is operated as a scientific partnership among Caltech, the University of California and NASA. The Keck Observatory was made possible by the generous financial support of the W.\,M.\ Keck Foundation.  This work is based in part on observations from the Legacy Science Program, made with the \textit{Spitzer Space Telescope}, which is operated by JPL, Caltech under a contract with NASA; and support for this work was provided by NASA through an award issued by JPL/Caltech. This research has also made use of the NASA/IPAC Extragalactic Database (NED) which is operated by JPL/Caltech, under contract with NASA.  Ned Wright's Javascript Cosmology Calculator was also used in preparing this paper \citep{Wright06}.

\setlength{\bibhang}{2.0em}

\end{document}